\documentclass[twocolumn,showpacs,preprintnumbers,amsmath,amssymb]{revtex4}

\usepackage{graphicx}
\usepackage{dcolumn}
\usepackage{bm}

\def\gev{\, \mbox{GeV}}

\def\del{\partial}
\def\Journal#1#2#3#4{{#1} {\bf #2}, #3 (#4)}

\def\NPB{{\em Nucl. Phys.} B}
\def\PLB{{\em Phys. Lett.}  B}
\def\PRL{\em Phys. Rev. Lett.}
\def\PRD{{\em Phys. Rev.} D}

\def\beq{\begin{equation}}
\def\eeq{\end{equation}}
\def\bea{\begin{eqnarray}}
\def\eea{\end{eqnarray}}

\begin{document}

\preprint{Freiburg-PHENO-2010-017.}

\title{Limits on a strongly interacting Higgs sector
\footnote{for Chris Quigg's 65th birthday}}

\author{J.~J.~van~der~Bij}
\affiliation{
Insitut f\"ur Physik, Albert-Ludwigs Universit\"at Freiburg \\
H. Herderstr. 3, 79104 Freiburg i.B., Deutschland
}

\date{\today}

\begin{abstract}
Using the classical argument about tree level unitarity breakdown in combination
with the precision electroweak data, it is shown, that if part
of the Higgs sector is strongly interacting, this part is small and is
out of range of the LHC.
\end{abstract}

\pacs{14.80.Ec, 11.80.Et, 12.15.Lk }
\maketitle

\section{Introduction}

Recently some renewed interest\cite{pomarol} in the possibility of a strongly interacting 
light Higgs 
sector has appeared. It was proposed to parametrize the effects of the strong
sector by anomalous couplings arising in the form of higher dimensional operators
at low energy.  Of course these operators should come from
an ultraviolet completion of the theory that gives rise to these effects.
The most important operators acting only on the Higgs field itself are:
\begin{equation}
{\cal O}_1 = \del_{\mu} (\Phi^{\dagger}\Phi)\del_{\mu} (\Phi^{\dagger}\Phi) .
\end{equation}
\begin{equation}
{\cal O}_2 = (\Phi^{\dagger}\Phi)^3 .
\end{equation}
These operators are automatically invariant under the custodial
$SU_L(2)\times SU_R(2)$ symmetry, which is highly desirable on phenomenological grounds.
The phenomenological effects are as follows.
After a rescaling of the fields ${\cal O}_1$ gives rise to anomalous Higgs boson couplings,
namely every coupling of standard model particles to the Higgs field is
multiplied with a common factor. Furthermore ${\cal O}_2$ gives rise 
to a change in the  Higgs selfcouplings. Measuring ${\cal O}_2$ would amount to measuring
the Higgs self-coupling, which is notoriously difficult at the LHC.
In precision tests at LEP, the effects of ${\cal O}_2$ appear only at the 
two-loop level and happen to be actually finite\cite{vdbij}, even though the full theory
is non-renormalizable. The effect is however very small.
In this paper we focus on ${\cal O}_1$, since the presence of this operator affects
the precision electroweak variables at the one-loop level and can therefore be
constrained more easily. Actually the theory with this operator is non-renormalizable
and this shows up already at the one-loop level in the electroweak
precision tests. The relevant corrections are logarithmically divergent. In order to realistically constrain
the theory one therefore has to start with an ultraviolet completion, that
gives rise to this operator at low energy. 

\section{The Hill model}

Such a completion is given by  the Hill model\cite{hill}, which is actually the simplest possible 
renormalizable extension of the standard model, having only two extra
parameters.
The Hill model is described by the following Lagrangian:
\begin{eqnarray}
{\cal L} =&& -\frac{1}{2}(D_{\mu} \Phi)^{\dagger}(D_{\mu} \Phi) 
-\frac {1}{2}(\partial_{\mu} H)^2 \\
&&- \frac {\lambda_0}{8} 
(\Phi^{\dagger} \Phi -f_0^2)^2  
-\frac {\lambda_1}{8}(2f_1 H-\Phi^{\dagger}\Phi)^2 .
\end{eqnarray}
Working in the unitary gauge one writes $\Phi^{\dagger}=(\sigma,0)$,
where the $\sigma$-field is the physical standard model Higgs field.
Both the SM Higgs field $\sigma$ and the Hill field $H$ receive vacuum expectation
values and one ends up with a two-by-two mass matrix to diagonalize, thereby
ending with two masses $m_-$ and $m_+$ and a mixing angle $\alpha$. There are two
equivalent ways to describe this situation. One is to say that one has two Higgs
fields with reduced couplings $g$ to standard model particles:
\begin{equation}
g_-= g_{SM} \cos(\alpha), \qquad g_+= g_{SM} \sin(\alpha) .
\end{equation} 
The standard model would correspond to $\alpha=0$ with the light Higgs
the standard model Higgs.
The other way, which has some practical advantages is not to diagonalize
the propagator, but simply keep the $\sigma - \sigma$ propagator explicitely.
One simply replaces the standard model Higgs propagator,
in all calculations of experimental cross section,
 by:
\begin{equation}
D_{\sigma\sigma} (k^2) = \cos^2(\alpha)/(k^2 + m_-^2) + \sin^2(\alpha)/(k^2 + m_+^2).
\end{equation}
The generalization to an arbitrary set of fields $H_k$ is straightforward, one 
simply replaces the singlet-doublet interaction term by:
\begin{equation}
L_{H \Phi}= - \sum \frac {\lambda_k}{8}(2f_k H_k-\Phi^{\dagger}\Phi)^2 . 
\end{equation}
For a finite number of fields $H_k$ no essentially new aspects appear,
however dividing the Higgs signal over even a small number of peaks,
can make the detection of the Higgs at the LHC very challenging.
Having an infinite number of Higgs fields one can also make a continuum,
which would make the Higgs field effectively undetectable at the LHC.
A mini-review of this type of models is given in \cite{jochum}.
There may even be evidence at LEP2 that this possibility is realized in 
nature\cite{dilcher}.
For the purpose of this paper the precise form of  the Higgs propagator
is irrelevant. Important is that there is a light piece in the Higgs
sector, which is weakly interacting and a heavy piece that is strongly interacting.
For this purpose the simple Hill model is sufficient.

\section{Limits for a strongly interacting Higgs sector}

In order to determine whether the Higgs sector can become strongly interacting
we adapt the classical analysis of \cite{quigg1,quigg2} to our case.
The breakdown of tree level unitarity is used as a criterium for
the presence or absence of strong interactions.  
Studying partial wave unitarity  the adapted classical analysis from \cite{quigg1,quigg2}
gives  the limit:
\begin{equation}
 cos^2(\alpha) m_-^2 +sin^2(\alpha) m_+^2 \, \leq \, \frac{8\pi \sqrt{2}}{3 G_F}.
\end{equation}  
in order to have tree level unitarity.
Since we demand that the Higgs sector becomes strongly interacting at high
energies we demand that this bound is broken. For this to happen one needs a 
sufficiently large combination $m_+ sin(\alpha)$. However one cannot have
an arbitrarily large value here, since the correction to a typical electroweak
precision observable $\delta_{EW}$ behaves like:
\begin{equation}
\delta_{EW} \approx \log(m_-^2/m_Z^2) + sin^2(\alpha) \log(m_+^2/m_-^2).
\end{equation}
This must then be smaller than the limit for the standard model
\begin{equation}
\delta_{EW} \leq \log(m_{up}^2/m_Z^2).
\end{equation}
where $m_{up}$ is the upper limit for the Higgs boson mass.
From the electroweak working group we take $m_{up}= 157 \gev$
We define $x=m_+/m_-$ and $m_{min}$ the minimal allowed Higgs mass, which
we take to be $m_{min}=115 \gev$ from the direct search.
The expectation value $v$ of the Higgs field is given by
$v^2= G_F^{-1}/\sqrt{2} = (246 \gev)^2$.
One then derives
\begin{equation}
\frac {x-1}{\log(x)} \,\, \geq \,\, \frac{16 \pi v^2}{3 m_-^2 \, 
\log(m_{up}^2/m_-^2)}.
\end{equation}
Taking $m_-=m_{min}$ one finds the weakest limits.
With the above values one finds:
\begin{equation}
m_+\, \geq 3285 \, \gev .
\end{equation}
 and
\begin{equation}
sin^2(\alpha)\, \leq \, 0.093 .
\end{equation}
So the lowest energy where  one can find a strongly interacting part of the Higgs
sector is at $3285 \gev$  with a production cross section of only 9.3\%
of the one for a standard model Higgs field with the same mass. This is clearly out of
range for the capabilities of the LHC, since this heavy Higgs boson is produced too little
because of its high mass and the reduced coupling to the standard model
particles. Moreover it is also wide, so there is no clear signal above the
background. The above analysis is of course somewhat simplified and could be improved
in many ways, for instance improving the unitarity bound, applying more accurate
formulas for the electroweak tests etc. However such refinements will not
change the conclusion, that strong interactions can only play a very
small part in the Higgs propagator
in a very high energy region, that is  out of the range of the LHC
or any machine that is at present under consideration.

\begin{acknowledgements}
This work was supported by the BMBF and by the EU-network
HEPTOOLS. I thank Dr. O. Brein for a critical reading of the manuscript.
\end{acknowledgements}


\end{document}